\documentclass[iop]{emulateapj}
\usepackage{graphicx}
\usepackage{natbib}
\usepackage{comment}
\begin{document}

\title{Supergiant Shells and Molecular Cloud Formation in the LMC}
\author{J. R. Dawson\altaffilmark{1}}
\email{joanne.dawson@utas.edu.au}
\author{N. M. McClure-Griffiths\altaffilmark{2}}
\author{T. Wong\altaffilmark{3}}
\author{John M. Dickey\altaffilmark{1}}
\author{A. Hughes\altaffilmark{4}}
\author{Y. Fukui\altaffilmark{5}}
\author{A. Kawamura\altaffilmark{6}}
\altaffiltext{1}{School of Mathematics and Physics, University of Tasmania, Sandy Bay Campus, Churchill Avenue, Sandy Bay, TAS 7005}
\altaffiltext{2}{Australia Telescope National Facility, CSIRO Astronomy \& Space Science, Marsfield NSW 2122, Australia}
\altaffiltext{3}{Astronomy Department, University of Illinois, Urbana, IL 61801, USA}
\altaffiltext{4}{Max-Planck-Institut f\"ur Astronomie, K\"onigstuhl 17, D-69117, Heidelberg, Germany}
\altaffiltext{5}{Department of Physics and Astrophysics, Nagoya University, Chikusa-ku, Nagoya, Japan}
\altaffiltext{6}{National Astronomical Observatory of Japan, Tokyo 181-8588, Japan}

\begin{abstract}
We investigate the influence of large-scale stellar feedback on the formation of molecular clouds in the Large Magellanic Cloud (LMC). Examining the relationship between H{\sc i} and $^{12}$CO(J=1--0) in supergiant shells (SGSs), we find that the molecular fraction in the total volume occupied by SGSs is not enhanced with respect to the rest of the LMC disk. However, the majority of objects ($\sim70\%$ by mass) are more molecular than their local surroundings, implying that the presence of a supergiant shell does on average have a positive effect on the molecular gas fraction. Averaged over the full SGS sample, our results suggest that $\sim12$--$25\%$ of the molecular mass in supergiant shell systems was formed as a direct result of the stellar feedback that created the shells. This corresponds to $\sim4$--$11\%$ of the total molecular mass of the galaxy. These figures are an approximate lower limit to the total contribution of stellar feedback to molecular cloud formation in the LMC, and constitute one of the first quantitative measurements of feedback-triggered molecular cloud formation in a galactic system. 
\end{abstract}


\section{Introduction}

The formation of cold, dense molecular clouds from the atomic interstellar medium is a key process in the evolution of galaxies, and one that 
sets fundamental boundaries on star formation rates. While recent work has shown that the presence of molecules is not strictly necessary for ongoing star formation \citep{glover12}, molecular tracers nevertheless provide the best observational diagnostic of the coldest, densest phase of the ISM in all but the most metal-poor systems. Astrophysical drivers of dense (molecular) cloud formation in disk galaxies include global gravitational instabilities \citep{wada00,kim02,tasker09,bournaud10,elmegreen11}, the accumulation of matter in spiral shocks \citep{kim06,dobbs06,dobbs08,dobbs12}, and compression in large-scale expanding shells driven by stellar feedback \citep{hartmann01,elmegreen02b,ntormousi11}; in all cases aided by turbulent compression that enhances density on a range of scales \citep[e.g.][]{elmegreen02,glover07}. However, disentangling the relative contributions of these processes is usually not trivial, and the primary drivers -- as well as the details of the physics -- are still open to debate. 

The role of stellar feedback in molecular gas production is of particular interest, since it is key element of how star formation self-regulates. The cumulative energy input from stellar winds, supernovae and ionizing radiation 
is able to form molecular clouds via the accumulation, compression and cooling of the ambient ISM in giant (100$\sim$1000 pc) supershells around OB clusters \citep{mccray87}. From a theoretical perspective, gravitational \citep[e.g.][]{elmegreen02b, wunsch12}, fluid dynamical \citep[e.g.][]{maclow89,vishniac94}, and thermal \citep[e.g.][]{inoue09} instabilities help to concentrate and fragment the walls of expanding shells into dense clouds; and more broadly, supershells fall within the scope of systems covered by the colliding flows theory of molecular cloud formation, in which molecular gas is formed rapidly from the atomic medium in the 
interfaces of turbulent ISM flows \citep[e.g.][]{audit05,vazquez06,vazquez07,hennebelle08,heitsch08c,banerjee09,inoue12,clark12}. Parameter space considerations suggest that small, thermally-driven condensations will develop before shells are able to become unstable on the longer wavelengths associated with gravitational instability \citep{heitsch08b}. 
However, global gravitational contraction helps to reduce overall timescales \citep{heitsch08a}, and results suggest that the onset of self-gravity and the conversion to the molecular phase likely occur at roughly the same time \citep{hartmann01,vazquez07}. 

Despite theoretical advances, there is as yet no consensus on contribution of supershells to molecular cloud formation rates, 
and there is a strong need for data to constrain theory. Observationally, the association of molecular clouds with supershells has been reported in a number of studies \citep[e.g.][]{jung96,koo88,carpenter00,kim00,mcclure00,matsunaga01,yamaguchi01b,yamaguchi01,dawson08,dawson11a}, and has often been interpreted as evidence of the in-situ formation of molecular gas in shell walls \citep[e.g.][]{fukui99,kim00,matsunaga01,yamaguchi01}. 
However, while the presence of molecular clouds in shell walls is indeed suggestive, it is important to make the distinction between association and formation; something that generally proves challenging for individual objects. Furthermore, large-scale stellar feedback can also be destructive -- both in the initial ionization and dynamical disruption of birth clouds \citep[e.g.][]{dale11,dale12,walch12}, and in the shock-destruction of pre-existing molecular clouds in the passage of an expanding shell \citep[e.g.][]{klein94,pittard11,dawson11b}. 

From a galactic evolution perspective, the \textit{net effect} of large-scale stellar feedback on the molecular gas fraction is therefore of critical importance -- the question of whether a supershell ultimately causes an increase or a decrease in the quantity of molecular gas in the volume it occupies. One of the first studies to directly address this question was \citet{dawson11a}, who found evidence of a net increase in the molecular gas fraction in two Galactic supershells. However, while the close proximity of Milky Way objects affords some advantages, Galactic studies suffer from difficulties in reliably identifying supershells and their associated emission in a confused Galactic Plane. To explore the question of the global influence of large-scale stellar feedback on a galactic system, we now turn away from the Milky Way, and towards our nearest star-forming neighbor, the Large Magellanic Cloud (LMC).

In this paper we compare the molecular and atomic content of the LMC disk within and outside of supergiant shell (SGS) systems in order to investigate the influence of large-scale stellar feedback on the molecular gas fraction of the ISM. SGSs are the largest and most energetic shells in the LMC, and those which we might expect to have the most dramatic effect on the evolution of the interstellar medium. The LMC is an attractive target for this kind of study, due to its relative proximity ($d\sim50$ kpc), nearly face-on orientation \citep[$i\sim35^{\circ}$;][]{vandermarel01}, and the presence of a large population of supershells and shell-like structures in its gaseous disk \citep[e.g.][]{meaburn80,kim99,chu90}. Excellent spectral-line data exist at arcminute resolutions for both the atomic and molecular components of the ISM; the former in the H{\sc i} 21 cm line \citep{kim98,staveley03} and the latter in $^{12}$CO(J=1--0) \citep{fukui99b,mizuno01,fukui08}. 
In section \ref{datasets} we describe these datasets. Section \ref{shelldef} then outlines our method of selecting and defining supergiant shells, describes the basic analysis and its assumptions, and examines sources of uncertainty. Section \ref{results} describes our results, which are discussed further in section \ref{discussion}. We finally summarize our conclusions in section \ref{summary}.

\section{Datasets} 
\label{datasets}

\subsection{H{\sc i}}

The LMC was observed in the H{\sc i} 21 cm line with the Australia Telescope Compact Array (ATCA) and the Parkes 64 m telescope, both of which are operated by CSIRO Astronomy and Space Science (CASS). Details of the observations and data combination are given by \citet{kim98,kim03} and \citet{staveley03}. The combined datacube covers an area of $7.5^{\circ} \times 7.5^{\circ}$ at an effective resolution of $1\arcmin$ ($\sim15$ pc), and a $1\sigma$ brightness temperature sensitivity of 2.4 K in a 1.65  km s$^{-1}$ velocity channel. 
For consistency with the CO data, the cube was regridded to the kinematic local standard of rest (LSR) velocity frame and smoothed to a resolution of $2.6\arcmin$. The main H{\sc i} integrated intensity ($I_{\mathrm{HI}}$) image used in this work is integrated over the velocity range $175 < v_{LSR} < 375 $ km s$^{-1}$, which contains all emission in the disk.

\subsection{$^{12}$CO(J=1--0)}
\label{coobs}

An unbiassed survey of the LMC in the 115 GHz $^{12}$CO(J=1--0) line was carried out with the 4 m NANTEN telescope in Las Campanas, Chile, operated by Nagoya University. The survey covered a total area of $\sim30^{\circ}$ and consisted of $\sim$ 26,900 positions observed on a $2\arcmin$ grid, with an integration time of $\sim3$ minutes per pointing. This spacing slightly undersamples the main beam, which has HPBW $\sim2.6'$ ($\sim40$ pc). The mean $1\sigma$ brightness temperature sensitivity is $\sim0.06$ K in a 0.65 km s$^{-1}$ velocity channel, but varies between $\sim0.04$ and $\sim0.09$ K in different sub-regions of the map. Further details of the observing strategy and data reduction are given by \citet{fukui08}.

Following \citet{wong11}, we apply a \textit{smooth-and-mask} technique to extract real emission features from the data. This method is is chosen because it is effective in recovering weak emission features from the CO cube, but results in no final degradation in resolution. The datacube is first convolved in the spatial domain with a Gaussian of twice the width of the telescope beam, smoothing it to a resolution of $6\arcmin$. This suppresses noise which should be uncorrelated between beam areas, but picks out any signal which is spatially resolved by the beam. A mask is then generated from this cube which retains only voxels detected above the $3\sigma$ noise level. 
This 
mask is applied to the original unsmoothed cube, and the resulting emission integrated over the full cube velocity range of $200 < v_{LSR} < 305 $ km s$^{-1}$ to produce a map of integrated intensity, $I_{\mathrm{CO}}$. This velocity range differs slightly from that of H{\sc i}, but contains all detected CO emission in the LMC \citep{mizuno01}. 
The mean RMS noise of 0.06 K per 0.65 km s$^{-1}$ channel corresponds to an integrated intensity sensitivity limit of 0.07 K km s$^{-1}$ for a line detection across three contiguous channels. We perform a final cleaning step to suppress noise in the $I_{\mathrm{CO}}$ map by blanking pixels below $4\sigma$ = 0.28 K km s$^{-1}$. This results in a final map in which weak spectral features are well-recovered with no loss of resolution.

\begin{figure*}
\epsscale{1.0}
\plotone{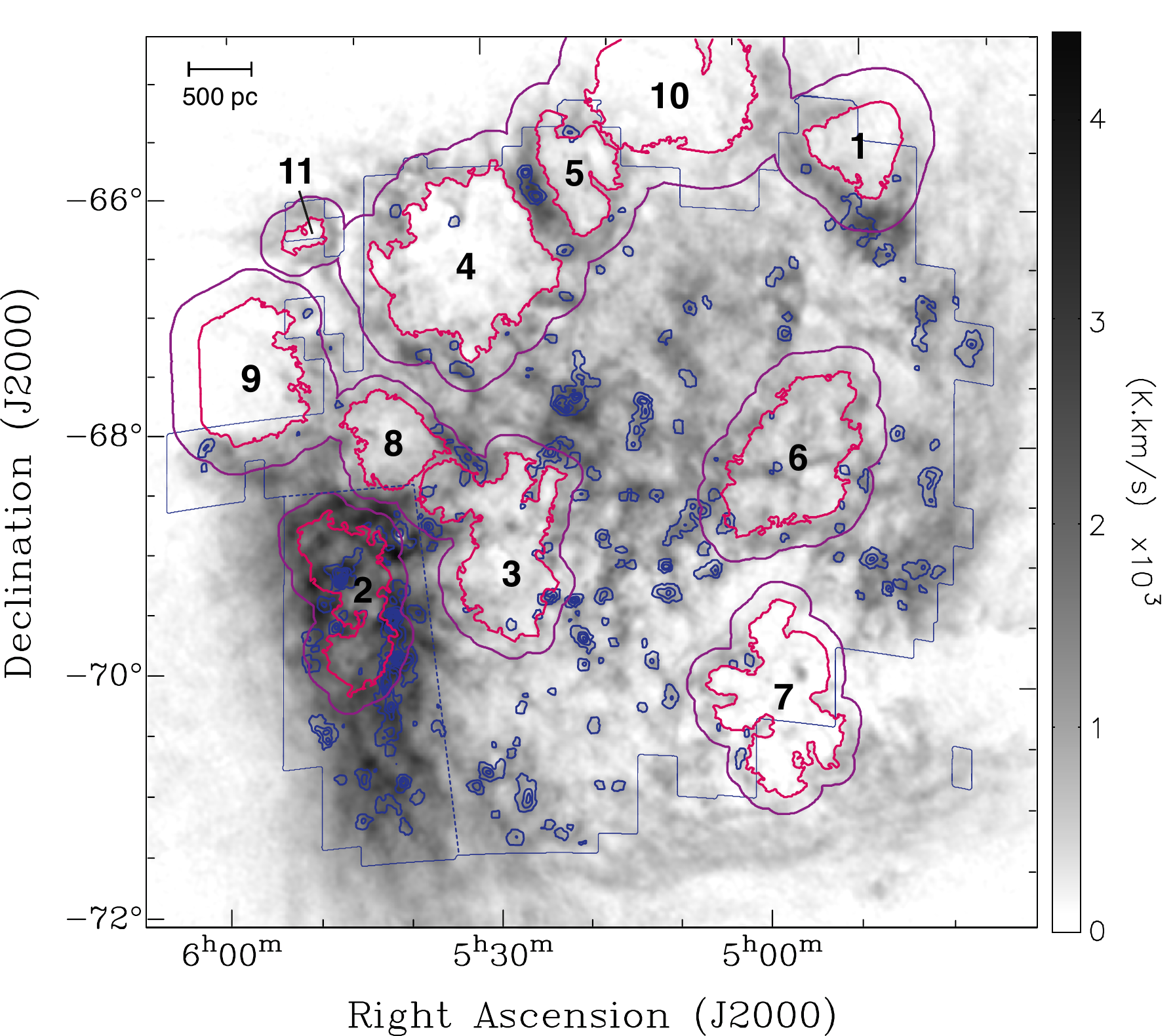}
\caption{Supergiant shells and shell complexes overlaid on integrated intensity map of the LMC. The greyscale image is H{\sc i} integrated over the velocity range $175 < v_{LSR} < 375 $ km s$^{-1}$ and smoothed to a resolution of $2.6\arcmin$. Blue contours are $^{12}$CO(J=1--0) integrated intensity, processed as described in \S\ref{coobs}, and integrated over the velocity range  $200 < v_{LSR} < 305 $ km s$^{-1}$. (Both images contain all emission detected in each tracer in the LMC). Contours start at 0.65 K km s$^{-1}$ and are incremented every 2.5 K km s$^{-1}$. The solid blue line marks the boundary of the region observed in CO. Dark pink lines trace the inner rims of the shell complexes and purple lines mark their outer boundaries, as described in \S\ref{shellmask}. Each shell complex is numbered and its constituent members listed in table \ref{table1}. Dotted blue lines enclose the region known as the southeastern H{\sc i} overdensity (SEHO).}
\label{fig:shellmap}
\end{figure*}

\section{Methodology}
\label{shelldef}

\begin{deluxetable*}{cclclccccccccccc}
\tabletypesize{\scriptsize}
\tablecolumns{15} 
\tablewidth{0pc} 
\tablecaption{SGS Complexes and Thresholding Parameters} 
\tablehead{ 
\colhead{SGS Complex} & \colhead{} & \multicolumn{3}{c}{Previous Listings} & \colhead{} &\multicolumn{7}{c}{Thresholding Parameters \tablenotemark{c}} & \colhead{} & \colhead{Shell \tablenotemark{c} (\arcmin)} & \colhead{$v_{exp}$ \tablenotemark{b,g}}\\ 
\cline{3-5} \cline{7-14} 
\colhead{} & \colhead{} & \colhead{H$\alpha$ \tablenotemark{a}} & \colhead{} & \colhead{H{\sc i} \tablenotemark{b}} & \colhead{} & \colhead{$\alpha$ \tablenotemark{d}} & \colhead{} & \colhead{$\delta$ \tablenotemark{d}} & \colhead{} & \colhead{$v_{LSR}$ range \tablenotemark{e}} & \colhead{} & \colhead{$T_b$ \tablenotemark{f}} & \colhead{} & \colhead{Thickness} & \colhead{km s$^{-1}$}\\
\colhead{} & \colhead{} & \colhead{} & \colhead{} & \colhead{} & \colhead{} & \colhead{(J2000)} & \colhead{} & \colhead{(J2000)} & \colhead{} & \colhead{km s$^{-1}$} & \colhead{} & \colhead{K} & \colhead{} & \colhead{} & \colhead{} 
}
\startdata
1 & & LMC1 & & SGS3 & & 04~59~00 & &-65~40~00 & & 270--285 & & 15 & & 17 & 15\\
2 & & LMC2 & & SGS20 & & 05~42~45 & &-69~55~00 & & 220--230 & & 35 & & 9 & 25\\
 & & LMC2 & & SGS19 & &05~44~45 & &-69~14~00 & & 260--270 & & 30 & &  & 25\\
3 & & LMC3 & & SGS8/12/13 & & 05~28~30 & &-69~17~00 & & 250--265 & & 20 & & 9 & 23/25\\
 & & LMC3 & & SGS15 & & 05~34~00 & &-68~44~00 & & 265--275 & & 20 & &  & 24\\
4 & & LMC4 & & SGS11/14 & & 05~31~45 & &-66~50~00 & & 275--300 & & 15 & & 15 & 36\\
5 & & LMC5 & & SGS7 & & 05~31~45 & &-66~50~00 & & 285--305 & & 20 & & 14 & 25/30\\
6 & & LMC6 & & SGS2 & & 04~59~15 & &-68~40~00 & & 250--270 & & 15 & & 13 & 17/20\\
 & & \nodata & & SGS5 & & 05~02~30 & &-68~25~00 & & 265--285 & & 15 & & & 20 \\
7 & & LMC8 & & SGS4 & & 05~01~00 & &-70~27~00 & & 220--240 & & 10 & & 9 & 23/15\\
8 & & \nodata & & SGS16/17/22 & & 05~39~00 & &-68~18~00 & & 275--280 & & 20 & & 7 & 18/26/? \tablenotemark{h} \\
9 & & \nodata & & SGS23 & & 05~51~30 & &-67~37~00 & & 275--285 & & 10 & & 15 & 23\\
10 & & \nodata & & SGS6 & & 05~14~30 & &-65~20~00 & & 280--290 & & 10 & & 19 & 18\\
11 & & \nodata & & SGS21 & & 05~45~00 & &-66~30~00 & & 280--290 & & 20 & & 13 & 19\\
\enddata
\label{threshtable}
\tablenotetext{a}{\citet{meaburn80}}
\tablenotetext{b}{\citet{kim99}}
\tablenotetext{c}{See \S\ref{shellmask}.}
\tablenotetext{d}{Approximate center position of thresholded void.}
\tablenotetext{e}{Velocity range over which emission is averaged prior to thresholding.}
\tablenotetext{f}{Threshold intensity level.}
\tablenotetext{g}{\citet{book08}}
\tablenotetext{h}{\citet{staveley03}}
\end{deluxetable*}

\subsection{Object Selection} 
\label{objects}

The disk of the LMC is rich in bubbles, shells, arcs and filaments, and numerous references to shells and supershells are present in the literature, identified in a variety of observational tracers \citep[e.g.][hereafter KDSB99]{davies76,meaburn80,chu90,kim99}. 
However, in general there is no one-to-one correspondence between different studies, with different tracers and methods favoring different classes of objects. A study on LMC shells must therefore give careful thought to its choice of sample.

The most comprehensive census of shells in the LMC disk is provided by the H{\sc i} 21 cm line, which probes the structure and dynamics of the atomic ISM. Supershells remain detectable in H{\sc i} long after their central energy sources have switched off, and objects defined in H{\sc i} should in theory comprise the most complete sample of shells and shell-like structures in the galaxy. The largest and most up-to-date catalogue of H{\sc i} supershells in the LMC was compiled by 
KDSB99. They define shells as coherent, bright-rimmed voids that are present in multiple velocity channels, and identify a total of 126 objects ranging in radius from $\sim50$ to $\sim700$ pc. 

In this work we consider only the `supergiant shells' (SGSs) of KDSB99 --  defined as those with radii greater than the scale height of the LMC gas disk ($z_g\sim180$ pc). \citet{book08} compare H{\sc i} and new H$\alpha$ data to examine in detail the SGSs from KDSB99 and the 9 similar objects from the classic H$\alpha$ catalogue of \citet{meaburn80}. Their analysis, with which we concur, ultimately rejects several SGSs as false detections, and refines the positions, velocities and extents of some others. Our final sample of supergiant shells is based on this revised selection, and contains 19/23 of the original KDSB99 catalogue. 
These are organized into 11 spatially distinct complexes (see \S\ref{shellmask}), which between them occupy $\sim40\%$ of the area of the main H{\sc i} disk. These complexes are shown in figure \ref{fig:shellmap} and are listed, along with their constituent shells, in table \ref{threshtable}. 

There are a number of reasons -- both scientific and pragmatic -- for working only with supergiant shells and excluding smaller objects. 
SGSs are a physically meaningful population of the largest and most energetic 
shells in the LMC, and are therefore expected to have the most dramatic effect on the ISM. Their regions of influence are large, giving them ample opportunity to accumulate the large quantities of material needed for molecular cloud formation. They are also
well-resolved at the resolution of the CO datacube, and far larger than a typical giant molecular cloud (GMC). This avoids ambiguity in the association of molecular gas with any given shell, and also avoids small-number problems that might arise from the association of only a few CO pixels with smaller objects.  
In addition, small/young supershells are expected to retain a correlation with the star-forming cloud complexes that birthed them, leading to difficulties in separating correlation and causality when interpreting their molecular gas fractions. 
SGSs, on the other hand, should be free of this bias, since their large size and advanced evolutionary state means that they have expanded far beyond their initial birth sites. 
In addition, SGSs are well studied, and multi-wavelength data exists for most objects, meaning that there is minimal ambiguity in their identification or origin. 
The same is not true of smaller H{\sc i} shells, many of which are likely less robust detections. 
For these reasons we restrict ourselves to exploring the influence of the largest and most obvious objects in the present work. 
However, we discuss the implications of excluding smaller shells in \S\ref{discussion1}.

Simulations demonstrate that it is possible to form giant dense-rimmed voids by gravity, turbulence and thermal instabilities alone, even without the inclusion of stellar feedback \citep[e.g.][]{wada00,dib05,dobbs11}. However, there is good evidence to suggest that the selected SGSs are genuine stellar feedback shells. Expansion is a strong signature of formation via a central energy source, and is confirmed in all of the constituent members of the eleven SGS complexes, 
in the form of H{\sc i} emission at velocities positive or negative of the systemic velocity  \citep[][see table \ref{threshtable}]{kim99,staveley03,book08}. \textit{ROSAT} observations by \citet{points01} find diffuse X-ray emission in six of the eleven complexes in our sample, interpreted as hot gas produced by stellar winds and supernova blasts. The exceptions are Complexes 6, 7, 9, 10 and 11. Of these, only Complex 6 was observed in their study; however, \citet{bomans96} do claim a detection for part of this complex in the same \textit{ROSAT} bands, so the literature is ambiguous. For Complex 7, \citet{chu11} claim that extended gamma ray emission in the direction of the shell arises from cosmic ray acceleration in SNR shocks, and young clusters close to the edges of the shell \citep{glatt10} are also directly implicated in expansion in its NE lobes \citep{book08}. Known clusters are similarly implicated in the dynamics of Complexes 1, 2, 3 and 4 \citep{book08,kawamura09,glatt10}. Of these, Complexes 2 and 4 bear special mention as objects for which a large body of research stretching back several decades makes a particularly robust and comprehensive case for 
a stellar origin \citep[see e.g.][]{caulet82,dopita85,bomans94,points99}. Little information exists on complexes 10 and 11. However, these objects cover very small areas in the observed region, and their impact on the results of this work is minimal (see also \S\ref{local}). 
It should be noted here that H$\alpha$ emission from shell walls is not a reliable indicator of a stellar origin, although it is sometimes treated as such. As pointed out by \citet{book08}, the ionizing radiation in H$\alpha$-bright SGSs frequently arises from young sources within shell walls and not from a central parent cluster. Similarly, an H$\alpha$-dark shell is one in which no recent ($\lesssim12$ Myr) star formation has occurred, but not one whose origins were necessarily non-stellar.

\subsection{Supergiant Shell Definition} 
\label{shellmask}

\begin{figure}
\plotone{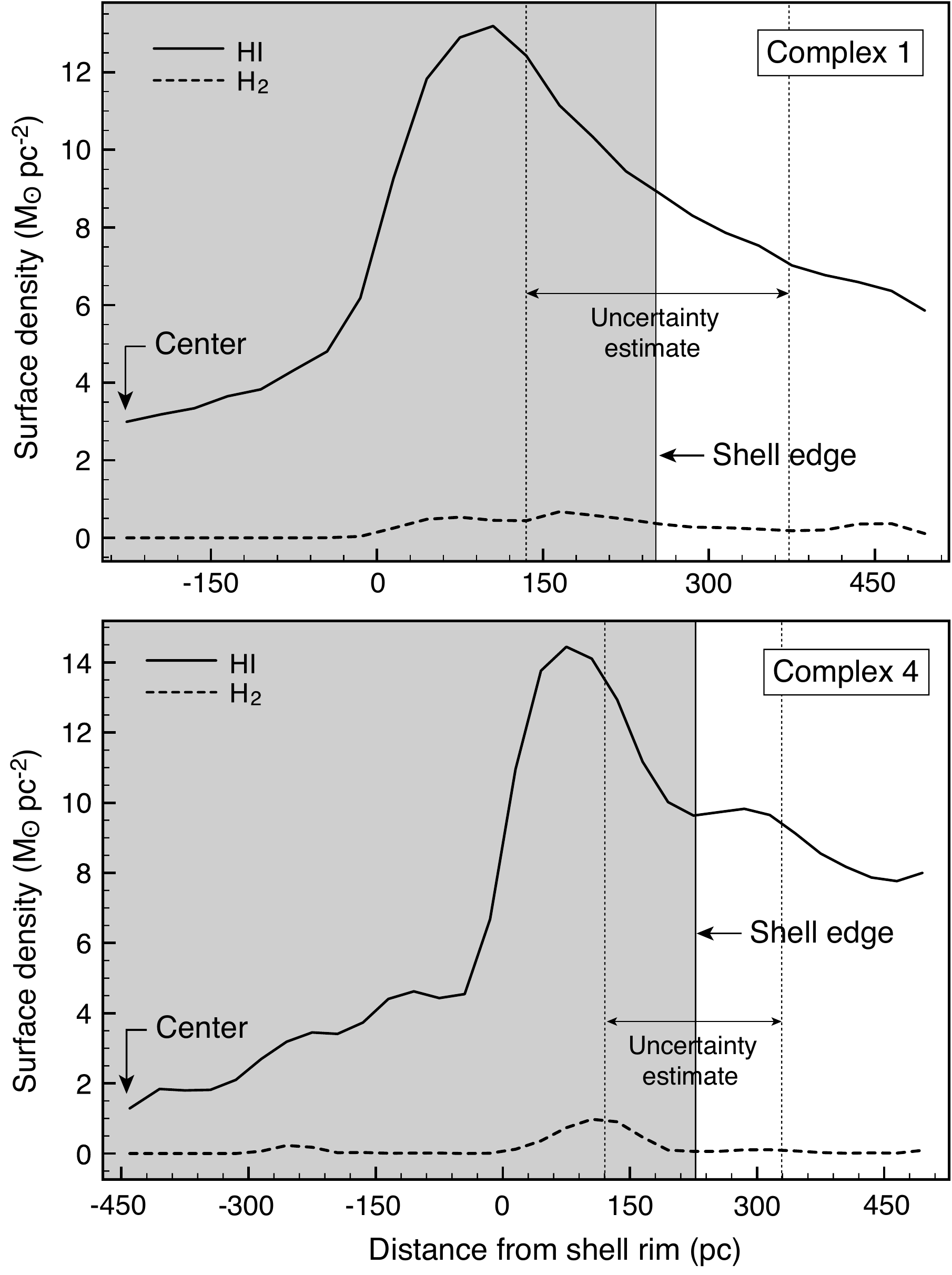}
\caption{Example of the definition of shell thickness for two SGS Complexes. The solid and dashed lines show variation of H{\sc i} and H$_2$ surface density with distance from the shell inner rim, calculated by summing emission within $2\arcmin$ bands incremented inwards and outwards from the thresholded inner rim of the SGS complexes. Units are $M_{\odot}$ pc$^{-2}$ with masses derived as described in \S\ref{approach}. The bright rim and evacuated void configurations can be clearly seen. Emission within the grey regions are defined as the SGS zones, whose edges is defined as described in \S\ref{shellmask}. Uncertainties are estimated as $\pm50\%$ of the shell thickness, and are marked with dotted lines. H$_2$ is included here for illustration, but is not used in defining the shells.}
\label{fig:lmc4plot}
\end{figure}

This work compares volumes affected by supergiant shells with volumes that are not, using two-dimensional velocity-integrated intensity maps summed over the full velocity range of the LMC. This 2D approach is well justified in the case of supergiant shells. Because SGSs are by definition larger than the H{\sc i} scale height of the LMC (KDSB99), they are assumed to have propagated through the full vertical extent of the gaseous disk to blow out on both sides of the plane. 
An SGS is therefore conceived as a cylindrical section of disk, all emission above and below of which is also related to the shell system. This mitigates the problems of line-of-sight contamination and shell curvature that would severely affect smaller objects. Emission at all velocities within the spatial extent of an SGS may therefore be assumed to be related to it. (The inclination of the LMC causes departures from this assumption, which we discuss below in \S\ref{uncertainty}) 


The shapes and extents of SGS regions are defined using a thresholding method, and pixel masks are generated that include the entirety of the bright shells and their evacuated voids. This method makes no ad-hoc assumptions about the morphology of a shell, instead allowing the configuration of the ISM to define its boundaries. 
SGSs are seen as rim-brightened voids in H{\sc i}, and are most clearly discerned at their systemic velocities. The following steps are performed to define individual SGSs and overlapping complexes:


\noindent1. A narrow-velocity H{\sc i} mean intensity map is produced by averaging the H{\sc i} brightness over the central few velocity channels of a supergiant shell, where the exact velocity range is chosen to include the entire bright rim of the shell over its full spatial extent ($\sim10$--$20$ km s$^{-1}$).\\ 
2. A contour is plotted at a level that picks out the shell's inner rim. Because H{\sc i} brightnesses in the LMC vary with both location and velocity, there is no `one size fits all' threshold level. The choice of threshold is made by eye, and is typically approximately half of the mean brightness of the shell walls.
Since the analysis in this work is insensitive to the exact positioning of the inner rim, this approach is acceptable. Similarly, since many shells are not completely enclosed rings, it is sometimes necessary to `fill in the gaps' in this rim by hand. Such gaps are usually small, and human judgement does not 
strongly influence the final size or shape.\\
3. All pixels within this inner rim are masked, and the masks for all shells are then combined. 
At this stage, overlapping objects are amalgamated into single complexes. 

Because many structures designated as SGSs by KDSB99 are either located completely within the spatial boundaries of larger shells, or are well connected to them in both space and velocity, it is not necessary to perform the above thresholding steps for every individual object in the KDSB99 catalogue. Where listed SGSs overlap significantly in both space and velocity, a single thresholding step at the appropriate velocity range recovers all separately catalogued features. Where there is full spatial overlap there is no need to threshold the smaller SGSs at all, since they occupy no unique pixels outside of their larger counterparts. Table \ref{threshtable} lists the approximate central positions, velocity ranges and threshold intensities used to generate the final SGS complexes.

At this stage the masks are imposed on the 2D H{\sc i} integrated intensity map, and contours are drawn at intervals of $2\arcmin$ incrementing both inwards and outwards from the rim of each shell complex. The H{\sc i} emission between each pair of contours is then summed to produce a plot of mean surface density vs distance from the complex rim. 
Example plots for Complexes 1 and 4 are shown in figure \ref{fig:lmc4plot}. Low-intensity voids and high-intensity walls are clearly seen, demonstrating how SGSs are a dominant component of LMC disk structure, even when emission is fully integrated in velocity. 
The outer edge of the wall is defined as the distance at which the mean H{\sc i} surface density 
has fallen 
to the first minimum, or in cases where the emission falls off smoothly with no subsequent rise (3/11 complexes), as the distance at which the mean surface density has decreased by half of the difference between the peak and the inner rim. It should be noted that the presence of a second peak outside the shell wall depends on the configuration of the surrounding ISM, and there are no differences observed in the properties of the SGS systems with and without this feature. 
The mask for each shell complex is then grown by this thickness. Thus, the final set of masks cover both the bright shells and evacuated voids of the LMC supergiant shells. 
The locations and outlines of the final SGS complexes are shown in figure \ref{fig:shellmap}. 

\subsection{The Metric: Molecular Fraction in SGS and `Background' Zones}
\label{approach}

The ISM in the volumes occupied by supergiant shells is compared to that outside them, to seek for evidence of molecular cloud formation in shell walls. We aim to answer the question of whether the passage of an SGS through the ISM causes the production of more H$_2$ than would have been created if no SGS had been present. The underlying assumption is that if the net effect of a supergiant shell is to produce more molecular matter in its swept-up walls than is destroyed by its initial passage through the ISM, then the volumes now occupied by SGSs should be more molecular than they were prior to the occurrence of the shell. 
The appropriate region over which to measure the molecular fraction is therefore the entire volume affected by a supergiant shell, including both the over-dense walls and the under-dense void (which in a 2D analysis will also contain emission from the front and back limbs of a shell). Selecting shell rims alone would preferentially sample only the highest H{\sc i} column densities, and would not probe the effect of SGSs on the full volumes of space that they influence.

\begin{figure*}
\plotone{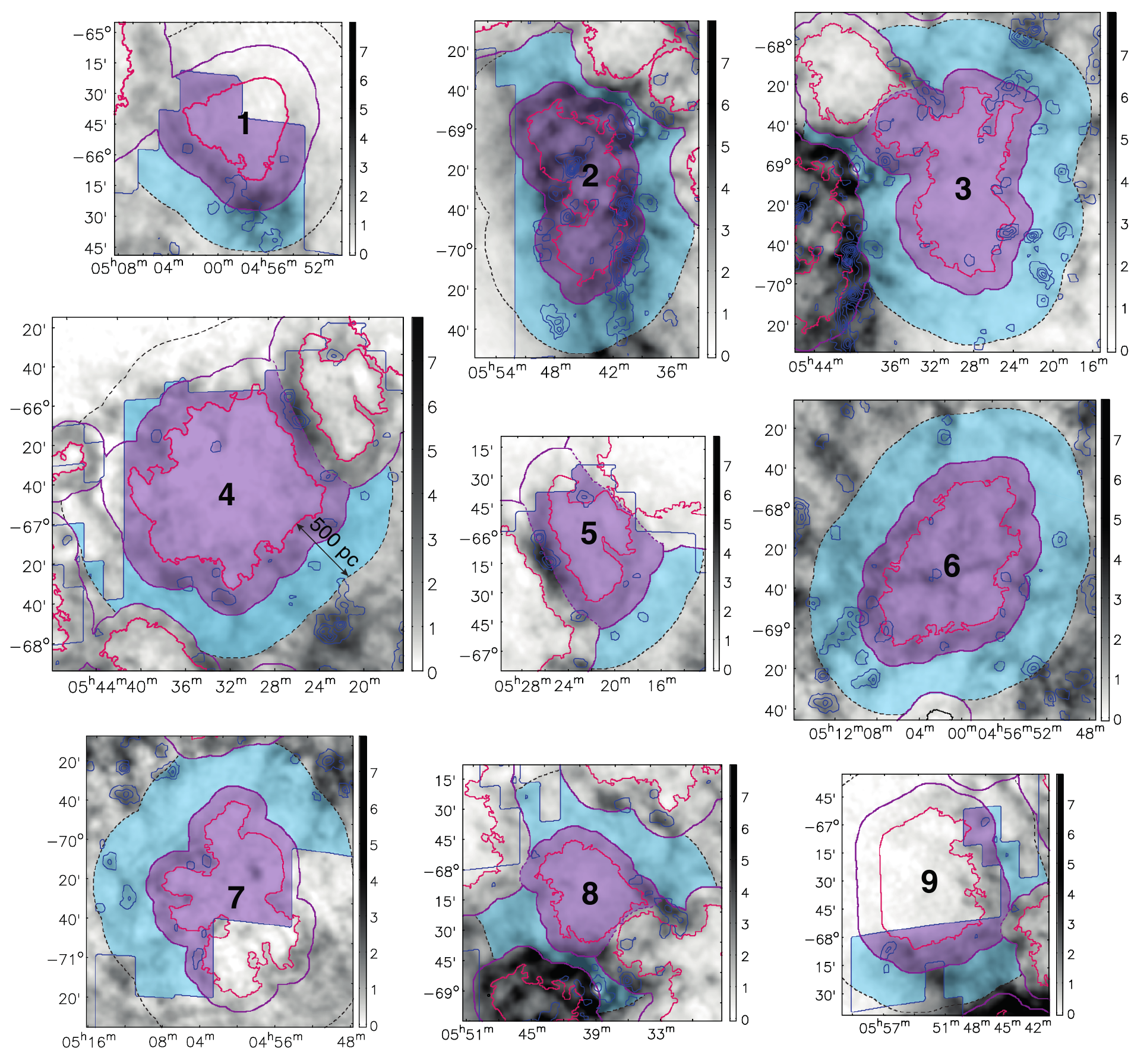}
\caption{SGS complexes and their local background zones, defined as described in \S\ref{approach}. Filled purple areas show the SGS zones and filled blue areas bordered by dashed lines shows the maximum widths of the local background zones (see \S\ref{local}). Spatial axes are right ascension and declination, and the color scale is in column density units of  $10^{21}$ cm$^{-2}$. All other lines and colors are as in Figure \ref{fig:shellmap}.}
\label{fig:bgzones}
\end{figure*}

In the absence of information on the ISM at past epochs, a comparison is made between SGSs and `background' (non-SGS) regions that serve as a proxy for the undisturbed medium. We consider two approaches. In \S\ref{global} SGS volumes are compared with the rest of the LMC disk. In this case SGSs between them occupy $45\%$ of the total observed area of the LMC, corresponding to 13.0 deg$^2$ (a projected area of $\sim11$ kpc$^2$), with a remaining 16.6 deg$^2$ ($\sim13$ kpc$^2$) designated as the background zone. In \S\ref{local} they are compared individually with their local surroundings, in order to minimize the impact of galaxy-scale variations in the molecular fraction. In the latter case, background zones are defined as bands of constant thickness around each shell complex, excluding pixels that are assigned to other shells. 
The minimum background zone width considered is $10\arcmin$ ($\sim150$ pc) from the shell edge, and the maximum considered is $33\arcmin$ ($\sim500$ pc) from the inner rim. 
$500$ pc represents an approximate limit at which emission is considered `local', roughly set by the distance at which the background zones of outer galaxy shells such as Complexes 4 and 8 begin to sample the very different environment of the dense central regions of the disk (see \S\ref{global}). 
SGS complexes are excluded from this individual analysis if either their shell or background zones do not include sufficient area in the CO datacube, defined as at least 0.25 square degrees of coverage ($\sim0.2$ kpc$^2$) in each zone. On this basis Complexes 10 and 11 (SGS6 and SGS21) are excluded. For this object-by-object analysis, pixels that belong to the swept-up wall of more than one complex (e.g. the bright ridge between Complexes 4 and 5) are assigned to the nearer of the two. Figure \ref{fig:bgzones} illustrates the SGS and maximum background zones for each complex. The surface areas occupied by SGSs and their partner background zones are given in table \ref{table1}.

We define two quantities, $f_{\mathrm{H}_2,\mathrm{SGS}}$ and $f_{\mathrm{H}_2,\mathrm{bg}}$,
where $f_{\mathrm{H}_2}$ is the molecular fraction of the ISM, given by
\begin{equation}
f_{\mathrm{H}_2} = \frac{M_{\mathrm{H}_2}}{M_{\mathrm{HI}} + M_{\mathrm{H}_2}},
\end{equation}

\noindent and the subscripts `SGS' and `bg' indicate measurements in supergiant shell volumes and background regions, respectively. $M_{\mathrm{HI}}$ and $M_{\mathrm{H}_2}$ are the atomic and molecular hydrogen masses measured from H{\sc i} and CO, and since the distance to all points in the LMC is assumed to be equal, $f_{\mathrm{H}_2}$ is simply related to column density as $f_{\mathrm{H}_2} = 2 N_{\mathrm{H}_2} / (N_{\mathrm{HI}} + 2N_{\mathrm{H}_2})$. $N_{\mathrm{H}_2}$ is calculated assuming an $I_{\mathrm{CO}}$ to $N_{\mathrm{H}_2}$ conversion factor of $3.0\times10^{20}$ cm$^{-2}$ K$^{-1}$ km$^{-1}$ s \citep[appropriate for the LMC;][]{leroy11}, and $M_{\mathrm{HI}}$ is assumed to be equal to $1.8\times10^{18}~I_{\mathrm{HI}}$ \citep{dickey90}; the exact result for optically thin gas. 

\subsection{Sources of Uncertainty}
\label{uncertainty}

The uncertainties inherent in estimating column densities from H{\sc i} and CO, and their potential effects on the observed relationship between $N_{\mathrm{HI}}$ and $N_{\mathrm{H}_2}$ in the LMC have been discussed extensively by \citet[][hereafter W09]{wong09}. For H{\sc i} the assumption of an optically thin gas will provide an underestimate of $N_{\mathrm{HI}}$ by a factor of $\tau/[1-e^{-\tau}]$, with absorption studies towards background continuum sources suggesting that this factor reaches a value of $\sim2.0$ towards the most opaque LMC sight lines \citep{dickey94,marx00}. For CO, a number of factors may affect the accuracy of the $N_{\mathrm{H}_2}$ conversion, particularly on the scales of individual clouds \citep[e.g.][]{shetty11a,shetty11b,genzel12,glover12}. 
These uncertainties are difficult to quantify. However, their impact on $f_{\mathrm{H}_2}$ is mitigated when summing over large regions of space, as we do here.  

The assumption that all emission along the line of sight to an SGS is genuinely part of the shell system implicitly assumes a face-on galaxy. For an inclination angle of $i\sim35^{\circ}$ \citep{vandermarel01} and an exponential H{\sc i} disk with a scale height of $\sim180$ pc \citep[KDSB99;][]{padoan01}, $\sim50\%$ of the H{\sc i} mass is contained in a layer $\sim250$ pc thick. A given line of sight through the LMC therefore contains emission that is spread over a projected distance of $\sim250~\tan{i}\approx175$ pc in the plane of the galaxy, translating to an in-plane positional uncertainty of $\pm87$ pc. This will not strongly affect the shapes of the SGSs themselves. The thresholded inner rims identified in \S\ref{shellmask} reflect a real structural property of the ISM -- the void/rim configuration that is the signature of a swept-up shell. Where problems may arise is in the assigning of emission to either an SGS or background zone, with the spreading of emission within the plane along a given sight line 
inevitably resulting in some emission being falsely assigned. This is not an effect that can be adequately expressed as an uncertainty on the defined shell thickness, nor can it be well modeled even by considering the third dimension of velocity. (Even on the steepest parts of the rotation curve $\delta R = 90$ pc translates to a rotational velocity difference of only $\lesssim3$ km s$^{-1}$; \citealt{kim98,vandermarel02,olsen07}.) It is a fundamental limitation of an inclined system that cannot be corrected for without knowledge of the 3D configuration of the gas. Fortunately, the molecular fraction in SGSs and their background zones is most sensitive to the placement of the scattered concentrations of CO, whose positional uncertainty is expected to be less severe. While the vertical distribution of molecular gas in the LMC is not known, CO scale heights are typically between two and several times smaller than H{\sc i}, both in the Milky Way \citep[e.g.][]{dickey90,malhotra94,stark05} and elsewhere \citep[e.g.][]{sancisi79,scoville93}. For an exponential CO scale height of $\sim90$ pc, the projected in-plane positional uncertainty on CO clouds is only $\sim\pm43$ pc -- approximately one resolution element. It is also worth noting that some estimates place the characteristic thickness of the LMC gaseous disk even smaller, at $\sim100$ pc for the H{\sc i} layer \citep{elmegreen01}, which would further reduce this uncertainty.

Our shell definition method relies on judging the end point of the increase in brightness associated with a SGS wall in the 2D image. Uncertainties associated with this step may be roughly quantified by a simple error estimate on the shell widths. This is implemented in the analysis in \S\ref{global} and \S\ref{local}, by varying the defined thickness of each shell by $\pm~50\%$ of its nominal value. A related concern is shell `leakage'. In a highly structured ISM, wind and supernova feedback may leak beyond the edges of a dense shell and out into the surrounding medium \citep[e.g.][]{chen00,walch12}. However, 
this is not likely to be a major issue in the present work. We are concerned with identifying regions where the ISM has been swept-up into dense structures by large-scale stellar feedback. Our method of defining SGSs directly recovers this behavior, by returning the edges of dense walls, making no ad-hoc assumptions about the morphology of a shell. 
If there is material flowing through breaks in shell walls, it will either be sufficiently energetic (and/or sufficiently confined) to excavate a new dense-walled cavity, or it will dissipate without strongly affecting the configuration of the neutral ISM. In the former case our method will accurately account for it, provided the new cavity is well joined to the original shell. Indeed, this kind of phenomenon can be seen in the case of Complex 8, which has several lobes extending from the main body of the SGS. Conversely, if the outflow is unable to influence the neutral gas sufficiently to form a void/wall configuration then 
its influence on the molecular cloud formation process is almost certainly negligible, and we are unconcerned with it in this work. We are therefore confidant that our shell definition method recovers the relevant material for our analysis.

\begin{figure}
\epsscale{1.1}
\plotone{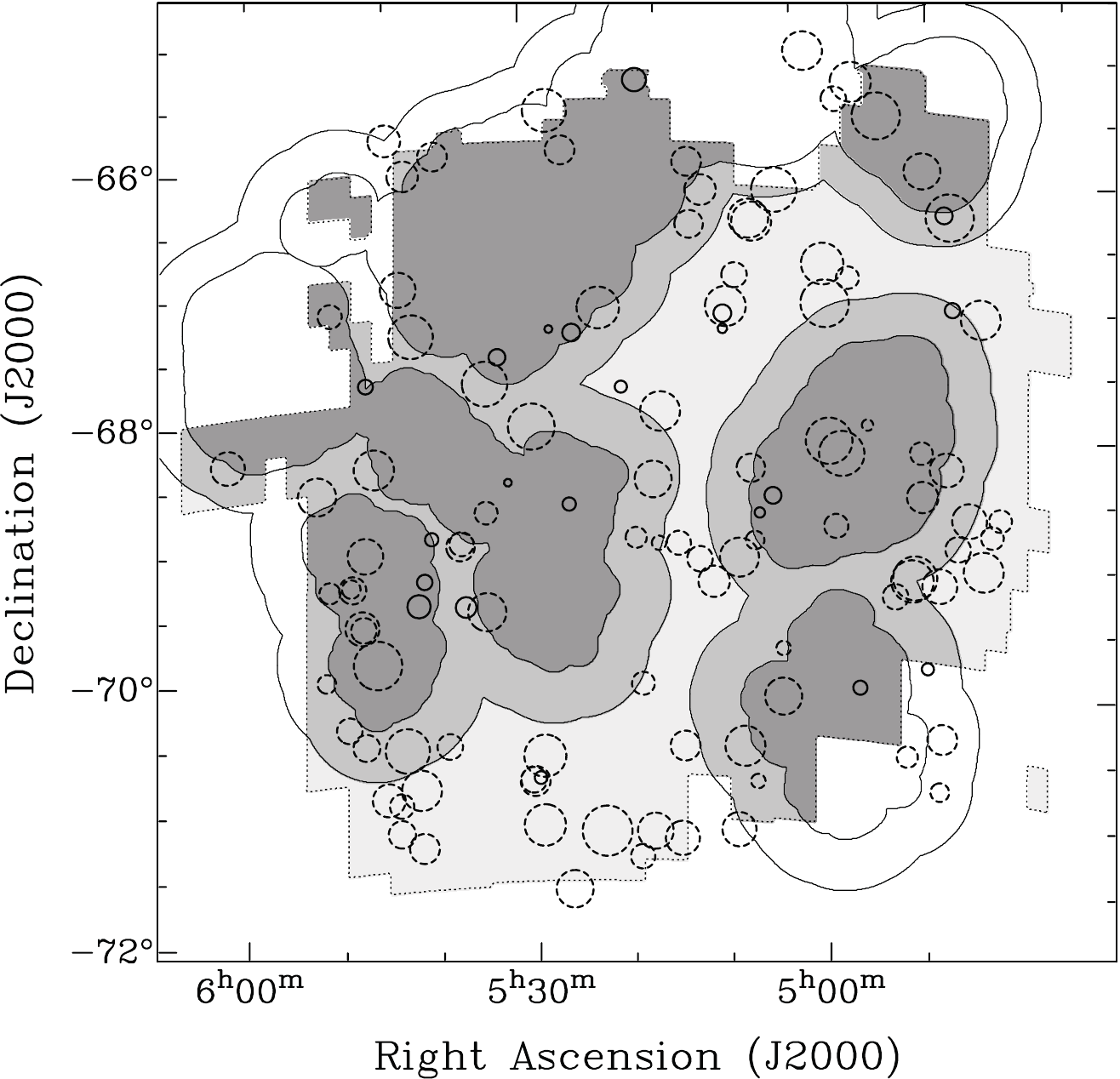}
\caption{Placement of smaller LMC supershells relative to SGSs and background zones. Dashed circles are the H{\sc i} `giant shells' of \citet{kim99}. Solid circles are H{\sc ii} regions that have been identified as superbubbles in the literature 
\citep{chu95,oey96,dunne01}. Shaded dark grey areas show the SGS zones used in the analysis, with shell widths defined as in table \ref{threshtable}. Shaded medium grey areas mark the maximum extent of the local background zones used in \S\ref{local}. The global background used in \S\ref{global} corresponds to the shaded light grey and shaded medium grey zones. 
The thin dotted line marks the boundary of the region observed in CO. 
} 
\label{fig:smallshells}
\end{figure}

The presence of smaller shells within the defined SGS and background zones may also potentially affect the analysis. If such shells significantly affect the molecular gas fraction, 
then a systematic overabundance of these objects in either SGS or background zones would bias our result. 
Figure \ref{fig:smallshells} overplots the H{\sc i} supershells of KDSB99 on a map of the SGS and local background zones, 
with H{\sc ii} regions identified in the literature as superbubbles also plotted. 
While it is possible that some of these H{\sc i} objects are not genuine feedback structures, they nevertheless provide a rough estimate of the distribution and abundance of smaller shells in the LMC disk. These shells cover very similar fractions of the total surface areas of the SGS and background zones. $0.19\pm0.01$ of the SGS zones, $0.18\pm0.01$ of the remainder of the LMC disk and $0.17\pm0.01$ of the aggregate local background zones are occupied by small shells, where the uncertainties reflect the values over the full range of shell and background zone widths used in the analysis. 
This suggests that to the first order, smaller shells may be assumed to contribute approximately equally to $f_{\mathrm{H}_2}$ in both SGS and background zones. Their presence is therefore unlikely to cause significant bias in the measurement of the effect of SGSs on the molecular fraction. 
The decision not to explicitly include smaller shells in the analysis does impact on the wider astrophysical interpretation of our results, however, which we discuss in \S\ref{discussion1}.

Finally, the simple method of defining local background zones 
may lead to a possible systematic bias in the analysis of individual shells. Many SGS complexes are located in the outer regions of the disk. However NANTEN CO coverage is limited at large galactocentric radii, affecting our ability to define background zones on all sides. As a result, background zones tend to preferentially sample material at smaller radii than their paired SGS zones. Since there is generally very little CO in the outer regions of galaxy disks \cite[whereas H{\sc i} remains strongly detectable, e.g.][]{wong02}, such an effect (if present) would result in values of $f_{\mathrm{H}_2,\mathrm{bg}}$ that are erroneously high, biassing the the analysis against a positive result. 

\section{Results}
\label{results}

\subsection{Global Molecular Gas Fraction}
\label{global}

We sum the H{\sc i} and CO emission within all SGS zones to derive a global molecular fraction for supergiant shells, and compare this to the global molecular fraction in the rest of the LMC disk. We find that the total molecular fraction in the regions occupied by SGSs is $f_{\mathrm{H}_2,\mathrm{SGS}}=0.054\pm0.004$. The equivalent quantity for the remainder of the observed area is $f_{\mathrm{H}_2,\mathrm{bg}}=0.055\pm0.002$. Here, uncertainties are estimated by varying the shell boundaries by $\pm50\%$ of the defined thicknesses, as given in table \ref{threshtable}.


The majority of the molecular gas in the LMC is located in one of two large 
regions 
in which $f_{\mathrm{H}_2}$ is enhanced with respect to the global average, and whose origins appear to be related to the global structure of the LMC disk. The closeness of $f_{\mathrm{H}_2,\mathrm{SGS}}$ and $f_{\mathrm{H}_2,\mathrm{bg}}$ 
is in fact somewhat serendipitous; arising from the fact that these regions of high 
molecular fraction are by coincidence equally represented in both SGS and background zones. 
The first of these CO-rich zones is the bright ridge-like feature sometimes known as the South-Eastern HI Overdensity (SEHO), which extends for $\sim3$ kpc along the South Eastern edge of the disk. This region is subject to tidal forces \citep[e.g.][]{staveley03} and ram-pressure effects from the LMC's passage through the Milky Way Halo \citep{deboer98}, and hosts $\sim1/3$ of the total molecular mass in the LMC. The SEHO is heavily sampled by SGS Complex 2. 
The second is the central portion of the disk, where large-scale gravitational instability \citep{yang07} and disk perturbations \citep{dottori96} arising from the influence of the old stellar bar likely combine to produce high gas densities. The arrangement of the supergiant shells is such that both shell and background zones sample approximately equal fractions of these high $f_{\mathrm{H}_2}$ regions, with the contribution of Complex 2, which includes most of the SEHO, offset by the lack of SGSs in the central disk. The mean H{\sc i} column density is also enhanced in these two regions, in keeping with the results of W09, who find that 
the probability of a CO detection increases with increasing $N_{\mathrm{HI}}$. 

This implies that SGSs are not the dominant driver of molecular gas production in the LMC; the global structure of the disk is a better determinant of where molecular matter is formed. Nevertheless, it may be that SGSs make some small contribution to molecular cloud formation, that can be recovered by comparison with background zones local to the shells. 

\subsection{Molecular Gas Fractions for Individual Shell Complexes}
\label{local}

The impact of unrelated global variations in $f_{\mathrm{H}_2}$ can be minimized by defining background zones locally, so that each SGS is paired with a background zone that only contains material in its immediate vicinity. This kind of comparison has already been carried out successfully for selected objects in the Milky Way \citep{dawson11a}.

Local background zones are defined as described in \S\ref{approach} and shown in figure \ref{fig:bgzones}. The molecular fraction in each individual SGS is calculated and compared with that in its paired background zone. 
To quantify sensitivity to the choice of region, $f_{\mathrm{H}_2,\mathrm{SGS}}$ and $f_{\mathrm{H}_2,\mathrm{bg}}$ are computed for 
a grid of shell and background zone widths. As for the global case, shell boundaries are varied by $\pm50\%$ of the defined thicknesses, as given in table \ref{threshtable}. Background zone widths are varied between the limits described in \S\ref{approach} (beginning at $10\arcmin$ from the shell edge and ending at $33\arcmin$ from the inner rim). Both shell and background zones are incremented in $2\arcmin$ intervals.   

\begin{figure}
\epsscale{1.1}
\plotone{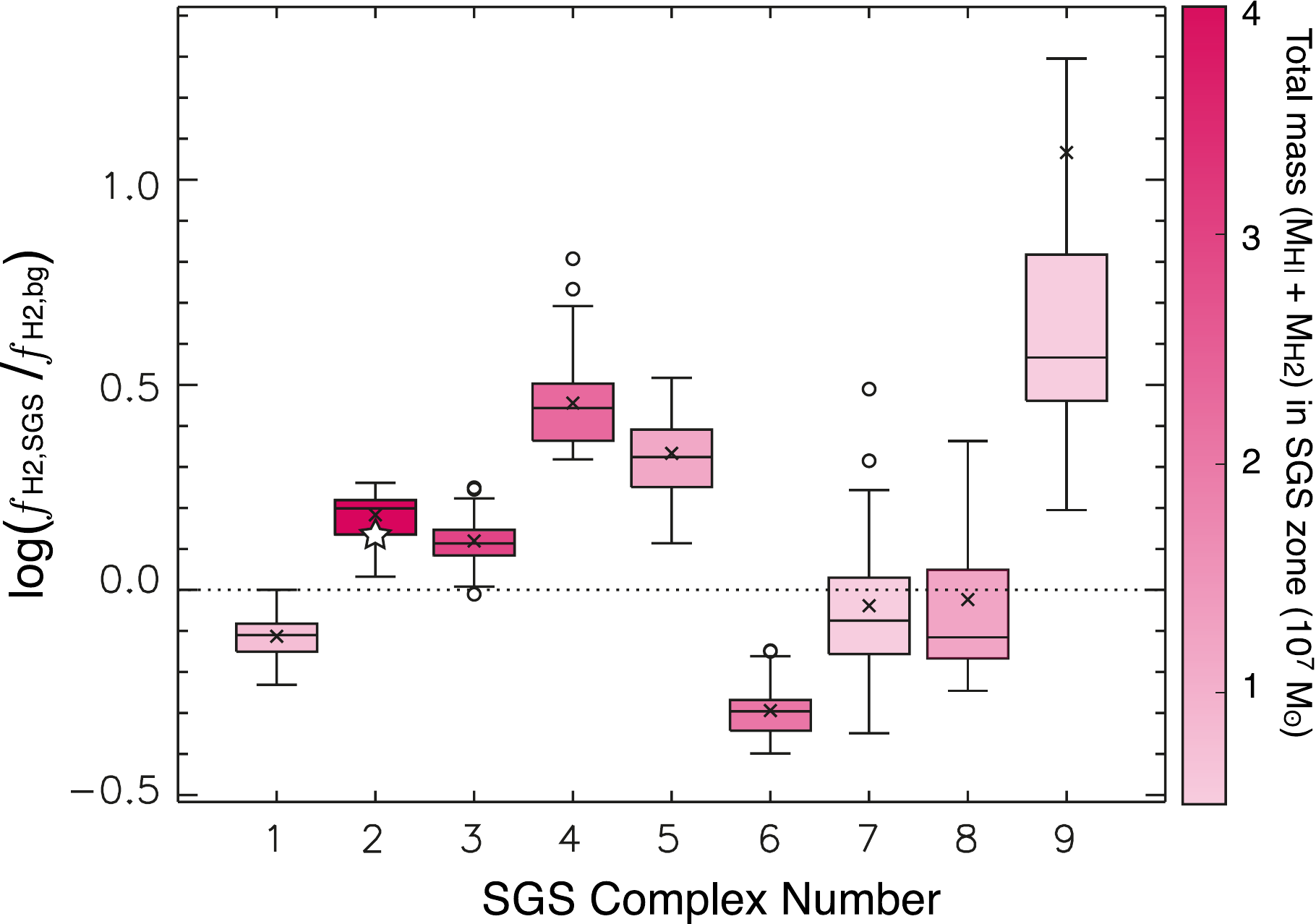}
\caption{Box plot of $\log(f_{\mathrm{H}_2,\mathrm{SGS}} / f_{\mathrm{H}_2,\mathrm{bg}})$ for individual SGS complexes and their local background zones. Each box illustrates the distribution of values obtained for a grid of background zones and shell widths, as described in \S\ref{local}. The central lines show the median values, the upper and lower bounds of the boxes show the upper and lower quartiles, and the whiskers extend out to the maximum and minimum values of the data, or to 1.5 interquartile range if there is data beyond this range. Means are plotted as crosses and outliers as circles. Complex 9 contains extreme high outliers arising from combinations where background zones contain almost no detected CO; these lie outside the vertical range of this plot and are not marked. Each complex is color-coded by the total mass, $M_{\mathrm{HI}} + M_{\mathrm{H}_2}$, contained in the SGS zone. The dotted line at $\log(f_{\mathrm{H}_2,\mathrm{SGS}} / f_{\mathrm{H}_2,\mathrm{bg}})=0$ marks the point at which the molecular fractions in the SGS and background zones are equal. The filled star marks the value obtained for Complex 2 (LMC2) when using the rest of the SEHO as the background zone (see text).}
\label{fig:boxplot}
\end{figure}

Figure \ref{fig:boxplot} shows a box plot of $\log(f_{\mathrm{H}_2,\mathrm{SGS}} / f_{\mathrm{H}_2,\mathrm{bg}})$ for each SGS complex, showing the maximum and minimum values, the lower and upper quartiles, and the mean and median results obtained over the full grid of shell and background sizes. 
We find that of the nine complexes, five show strong evidence for increased molecular fraction in the SGS volumes (Complexes 2, 3, 4, 5 and 9), two show good evidence of a decrease (Complexes 1 and 6) and two show no significant difference to within the $1\sigma$ uncertainties (Complexes 7 and 8). The median values of $\log(f_{\mathrm{H}_2,\mathrm{SGS}} / f_{\mathrm{H}_2,\mathrm{bg}})$ range from -0.29 to 0.57, and are listed in table \ref{table1}. Those complexes showing an apparent enhancement contain $\sim70\%$ of the total mass in the sample.  In this context it is interesting to note that the complex with the strongest apparent decrease (Complex 6) is dominated by one of the least robust SGS detections in the sample (see \S\ref{objects}). 


These numbers may be used to obtain a crude estimate of the contribution made by SGSs to the formation of molecular gas in the LMC. We compute the difference between the measured $M_{\mathrm{H}_2}$ in each SGS complex and an `expected' value derived from the median value of $f_{\mathrm{H}_2,\mathrm{bg}}$ -- assumed to represent the molecular fraction that would have been measured if no SGS had been present in the volume. The results for each complex are given in table \ref{table1}. In the case of Complex 2, 
an alternative approach is to use the remainder of the SEHO as the background zone, since there is evidence that the relationship between $N_{\mathrm{HI}}$ and $f_{\mathrm{H}_2}$ there may be different to the rest of the galaxy. 
This does not substantially change the results, but provides a slightly higher estimate of $f_{\mathrm{H}_2,\mathrm{bg}}$ (listed in the table and also marked on figure \ref{fig:boxplot}). Table \ref{table2} shows the total H{\sc i} and H$_2$ masses in the entire LMC disk (within the CO observed region), within SGS complexes, and the total mass estimated to have been created due to the influence of SGSs. Summing over the entire sample of shells, this estimate suggests that $\sim25\%$ of the molecular matter in SGS complexes may have been formed as a result of the stellar feedback that formed the shells. This is equivalent to $\sim11\%$ of the total molecular mass of the LMC. These figures fall to $\sim14\%$ and $\sim8\%$ when the background zone for Complex 2 is defined as the remainder of the SEHO. Although it is the contribution from Complex 2 that dominates these numbers, we note that a positive result ($\sim12\%$ of the CO in SGSs, corresponding to $\sim4\%$ of the total molecular mass in the disk) is still found when the SEHO is completely excluded from the analysis.

\begin{deluxetable*}{lcc}
\tablecolumns{3} 
\tablewidth{0pc} 
\tablecaption{Total H{\sc i} and H$_2$ masses in various regions of the LMC.} 
\tablehead{ 
\colhead{} & \colhead{$M_{\mathrm{HI}}/M_{\odot}$} & \colhead{$M_{\mathrm{H}_2}/M_{\odot}$}
}
\startdata
Whole LMC, where CO observed & $3.0~(2.1)\tablenotemark{b}\times10^8$ & $1.7~(1.1)\tablenotemark{b}\times10^7$\\
SGS complexes, where CO observed & $1.2~(0.8)\tablenotemark{b}\times10^8$ & $7.0~(3.3)\tablenotemark{b}\times10^6$\\
Estimated mass created due to SGSs \tablenotemark{a} & \nodata & $1.8~[1.4]\tablenotemark{c}(0.4)\tablenotemark{b}\times10^6$\\
\enddata
\tablenotetext{a}{See \S\ref{local}}
\tablenotetext{b}{Values in round brackets obtained when SEHO completely excluded from analysis.}
\tablenotetext{c}{Value in square brackets obtained when remainder of SEHO used as local background for Complex 2.}
\label{table2}
\end{deluxetable*}

We stress that these are crude estimates. Nevertheless, the results suggest that the formation of supergiant shells has a measurable positive effect on the molecular fraction of the LMC, albeit one that may be small compared to other drivers of molecular gas production in the galaxy. These numbers are also best considered as lower limits, as we will discuss below (\S\ref{discussion1}).


\subsection{CO Detection Fraction in Supergiant Shells}

\begin{figure}
\plotone{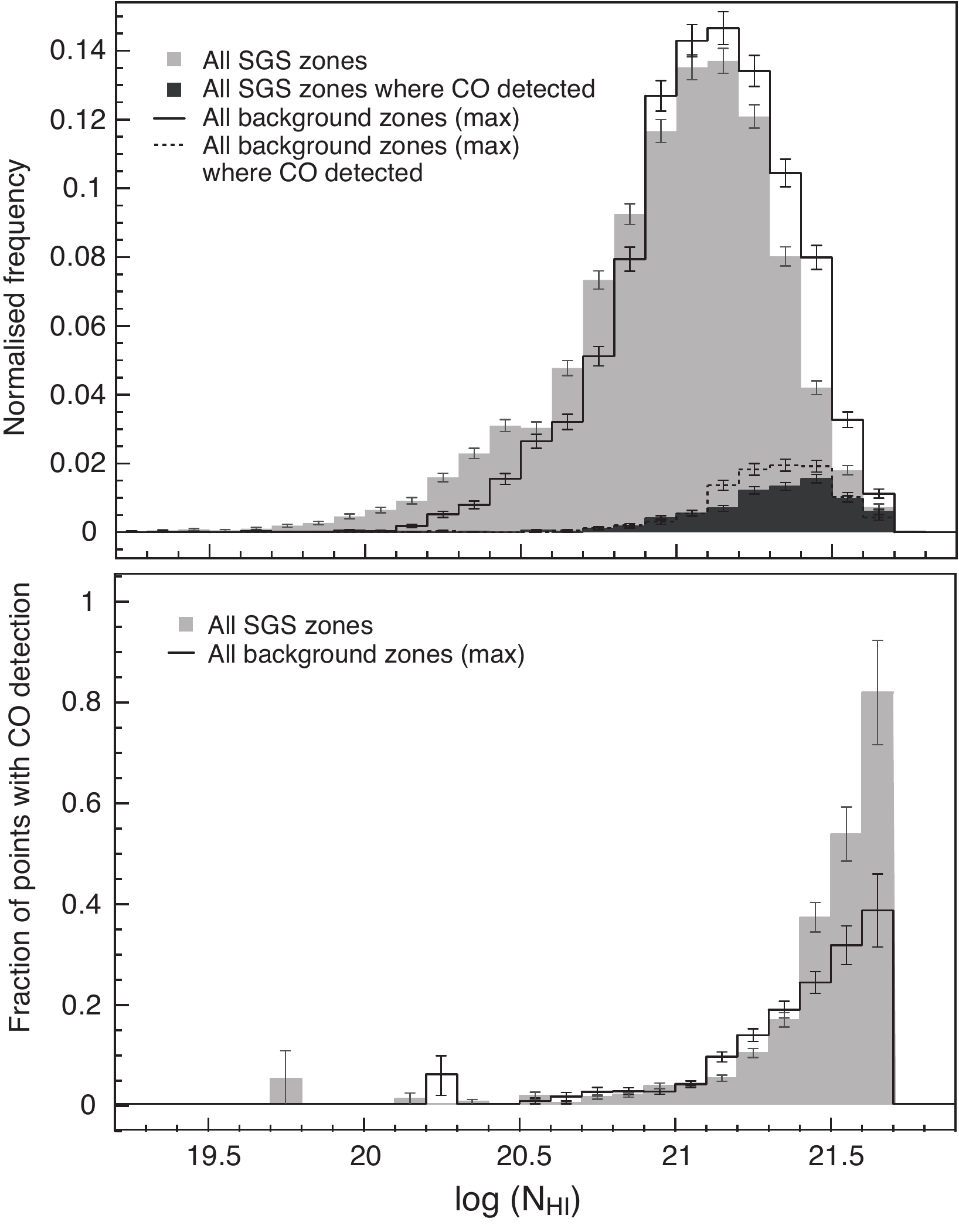}
\caption{Upper panel: $N_{\mathrm{HI}}$ histograms for the aggregate of SGS zones (light grey) and the aggregate of their local background zones (black line), excluding the SEHO. Dark shading and dotted lines show the distribution of pixels for which CO was detected in the SGS zones and background zones respectively. The results displayed here use the maximum background zone width (see \S\ref{local}), but the basic features of the distribution remain unchanged over the full range of widths tested. Lower panel: CO detection fraction histograms for aggregate of SGS zones (light grey) and the aggregate of their local background zones (black line), excluding the SEHO. Error bars are the scaled $1/\sqrt{N}$ counting errors. 
} 
\label{fig:hists}
\end{figure}

W09 find that the probability of detecting CO at any given position in the LMC disk increases with increasing $N_{\mathrm{HI}}$ (although it never reaches unity), and that where CO is present there is also a weak positive correlation between $N_\mathrm{HI}$ and $N_{\mathrm{H}_2}$. Given this, 
the enhanced $f_{\mathrm{H}_2}$ in SGS zones might be expected to arise from a tendency for them to possess a higher fraction of pixels at the largest H{\sc i} column densities. 
However, this does not appear to be the case. The top panel of figure \ref{fig:hists} shows H{\sc i} column density histograms for the SGS and local background zones (aggregated). The SEHO has been excluded, since the relationship between CO and H{\sc i} there is different to the rest of the disk (W09). The shells show a low column density tail corresponding to their evacuated voids, and their overall $N_{\mathrm{HI}}$ distribution is in fact skewed towards slightly lower values than the aggregate backgrounds. The figure illustrates the results for the largest background zones used in the analysis, but the key features of the distribution are insensitive to the choice of background zone width. The result is also not strongly driven by any particular complex and holds true on individual scales for all SGS and background zone pairs, with the exception of Complex 5.

Instead, the SGS volumes show an enhanced CO detection fraction at high $N_{\mathrm{HI}}$. 
The lower panel of figure \ref{fig:hists} shows the CO detection fraction as a function of H{\sc i} column density for the same aggregate SGS and background zones. The CO detection threshold for both samples is similar, at $\sim5\times10^{20}$ cm$^{-2}$. However, 
the SGS zones show a significantly enhanced likelihood of CO detection for $N_{\mathrm{HI}}>3\times10^{21}$ cm$^{-2}$. This behavior is mostly driven by several bright CO/H{\sc i} features in the dense walls of Complexes 3, 4 and 5 -- three of the five objects showing individually enhanced molecular fractions. Similarly, the lower detection fractions observed at high $N_{\mathrm{HI}}$ in the aggregate backgrounds are driven by a few patches of bright H{\sc i} with no CO. The small number of regions driving this trend, and the large scatter observed generally in the CO/H{\sc i} relationships throughout the LMC, urges caution in interpretation of this behavior. Nevertheless, it does suggest that the simple explanation of SGSs possessing higher H{\sc i} column densities and therefore more molecular gas does not satisfactorily explain the enhanced $f_{\mathrm{H}_2}$ in supergiant shells. 

\section{Discussion}
\label{discussion}

\subsection{The Role of Stellar Feedback in Molecular Cloud Formation}
\label{discussion1}

It has been proposed that while the feedback from massive stars plays a role in structuring the disk ISM \citep[e.g.][]{avillez01,joung09,dobbs11,hill12}, and 
is likely responsible for triggering a significant fraction of observed star formation on local scales \citep[e.g.][]{boss95,yamaguchi99,hosokawa06,dale07,leao09,deharveng10}, the initial production of dense (molecular) clumps and clouds on galactic scales is driven mainly by a combination of global gravitational instability and turbulence 
\citep[e.g.][]{wada01,elmegreen02,tasker09,bournaud10,maclow12}. Our results appear broadly consistent with this picture. While supergiant shells are clearly important in structuring the LMC disk, the results above suggest that they are not the dominant driver of the formation of molecular clouds. This is inferred both from the comparison of SGS volumes with the rest of the disk -- which shows that the global structure of the disk is a better determinant of where molecular matter is found -- and from the comparison of SGSs with their local environment -- which finds an enhancement in molecular fraction corresponding to only $\sim4$--$11\%$ of the total molecular mass of the LMC disk. 

An important caveat to the above interpretation is that the numbers derived in \S\ref{local} are (approximate) lower limits on the total contribution of stellar feedback to molecular cloud formation in the LMC. 
This is firstly due to the decision to ignore smaller supershells, which may also be playing a role in molecular cloud formation, but whose contribution is not directly investigated in this work. 
Secondly, our analysis is limited to a current ``snapshot'' of the LMC. 
If molecular clouds were formed and dissipated within the lifetimes of the shells -- as may be the case in objects formed through multiple generations of propagating star formation \citep[e.g.][]{dopita85} -- then such information is unrecoverable within the present work. 
The lifetime of a typical GMC in the LMC is roughly estimated by \citet{kawamura09} to be 20--30 Myr. Dynamical age estimates place most LMC supergiant shells at ages of $t\lesssim10$ Myr \citep{kim99}, although such estimates are necessarily crude. 
Estimates based on stellar population studies in individual SGSs point to somewhat older ages, typically 10--20 Myr \citep[e.g.][]{dopita85, points99, glatt10}. Thus the typical evolutionary timescales of SGSs do not exceed the lifetime of a typical GMC, and it seems unlikely that multiple generations of molecular clouds will form and die within the lifetime of a supergiant shell. However, if indeed GMC lifetimes are $\sim1.5$ times longer than the observable lifetime of an SGS, then clouds formed in shell walls will remain even after their parent shells have disappeared. Such clouds could fall in background zones, despite being feedback-formed.  This would bias our measurement towards erroneously low values for the fraction of molecular matter formed by SGS activity. 

\subsection{Final Comments: Dense Gas Tracers and Galaxy Types}

The quantity we have sought in this work is the fraction of the raw material for star formation that is produced in SGSs.  
At present, observational limitations mean that we are restricted to $^{12}$CO as a tracer, which cannot distinguish between diffuse molecular gas (that may be quiescent), and the very dense material that is directly implicated in star formation. While some studies have examined higher density tracers \citep[e.g.][]{johansson94,chin97,wong06,wang09}, these have so far been  restricted to a small number of objects. However, future observations will be able to target higher density gas tracers more widely, to directly probe the material that is most tightly correlated with the star formation rate \citep[see][and references within]{lada10,lada12}. \citet{seale12} have recently reported spatially resolved observations of several LMC star forming GMCs in the high-density tracers HCN and HCO$^+$. It will be interesting to see how the enhancement in $f_{\mathrm{H}_2}$ in supergiant shells is affected when only the densest molecular gas is considered. 

It will also be of interest to examine the relationship between supershells and molecular cloud formation in different classes of galaxies. The LMC is a dwarf irregular with low shear \citep{weidner10}, a larger H{\sc i} scale height than spirals \citep{brinks02} and only weak spiral structure. Low shear means shells can grow larger before they are deformed, high scale heights mean they can expand further before vertical blowout and depressurization, and the lack of a strong spiral potential mitigates against disruption by spiral arms. While Milky Way work suggests a significant quantity of molecular gas formation in two Galactic supershells \citep{dawson11a}, no systematic study of multiple shells has yet been performed to explore whether this behavior exists on large scales. Similarly, no quantitative comparison of CO and H{\sc i} in shells in other galaxies has been made to date.

\section{Summary and Conclusions}
\label{summary}

The large-scale stellar feedback from OB clusters is one of several astrophysical processes postulated to drive the formation of molecular clouds. We have used H{\sc i} and $^{12}$CO(J=1--0) data to examine the role of supergiant shells in the formation of molecular gas in the LMC, by comparing the molecular fraction, $f_{\mathrm{H}_2}$, in SGS volumes to that in undisturbed `background' regions. 

We have summed the H{\sc i} and CO emission over all SGS volumes to derive a global molecular fraction for supergiant shells, yielding $f_{\mathrm{H}_2,\mathrm{SGS}}=0.054\pm0.004$. This is identical (within the errors) to the molecular fraction in the remainder of the LMC disk. 
This suggests that supergiant shells are not a dominant driver of molecular cloud formation in the LMC. Indeed, the global structure of the LMC disk is generally a better determinant of where the highest molecular fractions are found. 

However when the impact of galaxy-scale variations in $f_{\mathrm{H}_2}$ is minimized, we find that supergiant shells have a positive effect on the molecular gas fraction in the volumes of space they occupy. This is measured by comparing SGSs with their local surroundings, which are a better proxy for the undisturbed medium of individual objects. 
In this local analysis the majority of SGSs (5 out of 9 complexes, equal to $\sim70\%$ by mass) exhibit molecular fractions that are significantly enhanced with respect to their local background regions. Averaged over the full population, our results imply that $\sim12$--$25\%$ of the molecular mass in supergiant shell systems was formed as a direct result of their action on the ISM. This corresponds to $\sim4$--$11\%$ of the total molecular mass of the LMC.

These figures are an approximate lower limit to the total contribution of stellar feedback to molecular cloud formation in the LMC, and constitute one of the first quantitative measurements of feedback-triggered molecular cloud formation in a galactic system. The results of this work are tentatively consistent with a scenario in which large-scale gravitational instabilities -- as well as other global dynamical processes such as tidal forces and ram pressure in the South Eastern regions -- are responsible for the majority of dense gas formation in the LMC, with stellar feedback making an important secondary contribution. However, a detailed analysis, including smaller feedback structures and considering quantitatively the relative observable lifetimes of molecular clouds and shells, should be undertaken before this conclusion can be argued strongly. 

\acknowledgements
We thank the anonymous referee, whose thorough reading of the manuscript considerably improved this work. J. Dawson thanks Simon Ellingsen, Stas Shabala and Jamie McCallum for fruitful discussions. The Australia Telescope Compact Array and Parkes Telescopes are part of the Australia Telescope which is funded by the Commonwealth of Australia for operation as a National Facility managed by CSIRO. The NANTEN project was based on a mutual agreement between Nagoya University and the Carnegie Institute of Washington, and its operation was made possible thanks to contributions from many companies and members of the Japanese public.

\bibliographystyle{apj}
\bibliography{lmcbib}

\begin{turnpage}
\begin{deluxetable*}{ccccccccccccc}
\tabletypesize{\scriptsize}
\tablecolumns{11} 
\tablewidth{0pc} 
\tablecaption{Analysis of Molecular Fractions in Individual SGS Complexes and Local Backgrounds} 
\tablehead{ 
\colhead{SGS Complex} & \colhead{} & \multicolumn{3}{c}{Molecular Fraction $f_{\mathrm{H}_2}$\tablenotemark{a}/$10^{-2}$} & \colhead{} & \colhead{$\log(f_{\mathrm{H}_2,\mathrm{SGS}} / f_{\mathrm{H}_2,\mathrm{bg}}$) \tablenotemark{d}} & \colhead{} & \colhead{$\delta M_{\mathrm{H}_2}/M_{\odot}$ \tablenotemark{e}} & \colhead{}  & \multicolumn{3}{c}{Area/deg$^{2}$ \tablenotemark{h}}\\ 
\cline{3-5} \cline{11-13} \\
\colhead{} & \colhead{} & \colhead{SGS Zone\tablenotemark{b}} & \colhead{} & \colhead{Local Background\tablenotemark{c}} & \colhead{} & \colhead{} & \colhead{} & \colhead{} & \colhead{} & \colhead{SGS Zone} & \colhead{} & \colhead{Local Background\tablenotemark{g}}
}
\startdata
1 & & $3.8\pm0.4$ & & $4.7\pm0.4$ & & -0.11 & & $-1.1\times10^5$ & & 0.64--1.10 & & 0.29--0.66\\
2 & & $8.7\pm0.2$ & & $5.6\pm0.2$ & & 0.20 [0.13] \tablenotemark{f} & & $1.5~[1.0]$ \tablenotemark{f}$\times10^6$ & & 1.14--1.76 & & 0.78--1.86 [1.95--2.56] \tablenotemark{f}\\
3 & & $6.3\pm0.2$ & & $4.8\pm0.4$ & & 0.12 & & $3.0\times10^5$ & & 1.62--2.45 & & 0.89--2.60\\
4 & & $3.1\pm0.2$ & & $1.2\pm0.2$ & & 0.44 & & $3.6\times10^5$ & & 2.28--3.40 & & 0.58--1.31\\
5 & & $5.1\pm0.3$ & & $2.6\pm0.6$ & & 0.32 & & $2.7\times10^5$ & & 0.78--1.36 & & 0.25--0.44\\
6 & & $2.8\pm0.4$ & & $5.4\pm0.3$ & & -0.29 & & $-5.1\times10^5$ & & 1.65--2.77 & & 0.92--2.70\\
7 & & $3.0\pm0.4$ & & $3.3\pm0.7$ & & -0.07 & & $-2.2\times10^4$ & & 1.05--1.51 & & 0.57--1.65\\
8 & & $1.0\pm0.1$ & & $1.2\pm0.4$ & & -0.12 & & $-2.5\times10^4$ & & 0.66--1.01 & & 0.39--0.86\\
9 & & $2.8\pm0.3$ & & $0.8\pm0.1$ & & 0.57 & & $1.1\times10^5$ & & 0.34--0.91 & & 0.25--0.78\\
10 & & \nodata & & \nodata & & \nodata & & \nodata & & \nodata \\
11 & &  \nodata & & \nodata & & \nodata & & \nodata & & \nodata\\
\enddata
\tablenotetext{a}{Molecular fraction $f_{\mathrm{H}_2}=M_{\mathrm{H}_2}/(M_{\mathrm{H}_2}+M_\mathrm{HI})$.}
\tablenotetext{b}{Value for standard shell thickness as listed in table \ref{threshtable}. Uncertainties obtained by varying shell thicknesses by $\pm~50\%$.}
\tablenotetext{c}{Mean and standard deviation obtained over full range of background zone widths, as described in \S\ref{approach}.}
\tablenotetext{d}{Median values as derived in \S\ref{local}.}
\tablenotetext{e}{Estimate of H$_2$ mass difference due to presence of SGS. (See \S\ref{local})} 
\tablenotetext{f}{Values in square brackets obtained when remainder of SEHO used as local background for Complex 2.}
\tablenotetext{g}{Area for maximum width of $33\arcmin$ from SGS inner rim.}
\tablenotetext{h}{Maximum and minimum values within full range of SGS and background zone widths, as described in \S\ref{approach}}
\label{table1}
\end{deluxetable*}
\clearpage
\end{turnpage}

\end{document}